\documentclass[twocolumn,showpacs,preprintnumbers,amsmath,amssymb,superscriptaddress]{revtex4-1}
\usepackage{graphicx}
\usepackage{dcolumn}
\usepackage{bm}
\usepackage {longtable}

\begin{document}

\preprint{}

\title{Cooper minima in the transitions from low-excited and Rydberg states of alkali-metal atoms}
\author{I.~I.~Beterov}
\email{beterov@isp.nsc.ru}
\affiliation{A.V.Rzhanov Institute of Semiconductor Physics SB RAS, Prospekt Lavrentieva 13, 630090 Novosibirsk, Russia}
\author{C.~W.~Mansell}
\affiliation{The Open University, Walton Hall, MK7 6AA, Milton Keynes, UK}

\author{E.~A.~Yakshina}
\affiliation{A.V.Rzhanov Institute of Semiconductor Physics SB RAS, Prospekt Lavrentieva 13, 630090 Novosibirsk, Russia}

\author{I.~I.~Ryabtsev}
\affiliation{A.V.Rzhanov Institute of Semiconductor Physics SB RAS, Prospekt Lavrentieva 13, 630090 Novosibirsk, Russia}

\author{D.~B.~Tretyakov}
\affiliation{A.V.Rzhanov Institute of Semiconductor Physics SB RAS, Prospekt Lavrentieva 13, 630090 Novosibirsk, Russia}

\author{V.~M.~Entin}
\affiliation{A.V.Rzhanov Institute of Semiconductor Physics SB RAS, Prospekt Lavrentieva 13, 630090 Novosibirsk, Russia}

\author{C.~MacCormick}
\affiliation{The Open University, Walton Hall, MK7 6AA, Milton Keynes, UK}

\author{M.~J.~Piotrowicz}
\affiliation{The Open University, Walton Hall, MK7 6AA, Milton Keynes, UK}
\affiliation{University of Wisconsin-Madison, 1150 University Avenue, Madison, Wisconsin 53706, USA}

\author{A.~Kowalczyk}

\affiliation{The Open University, Walton Hall, MK7 6AA, Milton Keynes, UK}
\author{S.~Bergamini}

\affiliation{The Open University, Walton Hall, MK7 6AA, Milton Keynes, UK}

\date{12 July 2012}

\begin{abstract}
The structure of the Cooper minima in the transition probabilities and
photoionization cross-sections for low-excited and Rydberg \textit{nS},
\textit{nP}, \textit{nD} and \textit{nF} states of alkali-metal atoms has
been studied using a Coulomb approximation and a quasiclassical model. The
range of applicability of the quasiclassical model has been defined from
comparison with available experimental and theoretical data on dipole
moments, oscillator strengths, and photoionization cross-sections. A new
Cooper minimum for transitions between rubidium Rydberg states has been
found.
\end{abstract}


\maketitle

\section{Introduction}

Spectral line series of alkali metal atoms display remarkable features with
prominent minima in the transition probabilities, emission oscillator
strengths or photoionization cross-sections~\cite{Bates, Aymar1976, Cooper1962, FanoCooper, Ditchburn1943, TheodosiouMin1980,  HudsonCarter1967,  Aymar1978,
 Manson, Hoogenraad1995, HoogenraadDipole, Duncan,
Petrov}. These minima arise from the cancellation of the radial integral for
some transitions, depending on the overlap between the wavefunctions of the
initial and final quantum states of the atoms~\cite{Bates, Aymar1976}, and are
well known as Cooper minima~\cite{Cooper1962, FanoCooper}. 

Observation of the Cooper minima  in the photoionization cross-sections and
in the transition probabilities of the discrete spectrum provides valuable
information about the electronic structure of the atoms. Minima in
photoionization cross sections were first found experimentally in Ref.~\cite{Ditchburn1943} and explained 20 years later by Cooper~\cite{Cooper1962}. The
sharp minima in the emission probabilities for some Rydberg states of
alkali-metal atoms were first discussed by Theodosiou~\cite{TheodosiouMin1980}.

The experimental investigation of Cooper minima can be used for the
verification of theoretical calculations of spectroscopic properties of
atoms and molecules. For example, experimentally measured photoionization
cross-sections for sodium ground state showing the Cooper minimum~\cite{HudsonCarter1967}, were in good agreement with the theoretical calculations
of Aymar~\cite{Aymar1978}, which confirmed the accuracy of the theoretical model.

Cooper minima in the discrete spectrum are revealed as a suppression of the
two-photon photoionization~\cite{Hoogenraad1995} or sharp decrease of the
emission oscillator strengths~\cite{TheodosiouMin1980}. The map of these minima
could be valuable for systematic studies of the processes which involve a
large number of transitions, such as calculation of lifetimes of Rydberg
atoms~\cite{BeterovLifetimes}, blackbody-radiation-induced photoionization rates~\cite{BBRIonization, BeterovNJP} or collisional ionization cross-sections of cold
atoms~\cite{Amthor2009}.

Photoionization of alkali-metal atoms recently attracted a lot of interest,
as it has taken a central role in experiments with cold atoms in
far-off-resonance traps~\cite{AdamsPhoto, ShafferCs}, in photoionization
spectroscopy~\cite{RbDroplets, Haq2010, Nadeem2011},   measurement of oscillator
strengths~\cite{Haq2010, Nadeem2011,Piotrowicz2011}, photoionization cross-sections and lifetimes of excited atoms~\cite{Haq2010, Nadeem2011, Piotrowicz2011, Fabry, LiOscStr, BaigNa, Gabbanini2006,BeterovLifetimes, Feng, Tate}, 
and photoionization of the Bose-Einstein condensate~\cite{BECion,AlkaliLasers}.

Rydberg atoms with large principal quantum numbers \textit{n}$\sim $50-100
recently received attention due to the progress achieved in experiments with
cold atomic samples. These samples are often prepared in optical
traps whose intense laser field could make the ionization lifetimes of
Rydberg atoms extremely short. However, for certain wavelengths, ionization
rates could be significantly reduced due to Cooper minima in the
photoionization cross-sections~\cite{Cooper1962}. Therefore it would be most
useful to exploit trapping schemes operated at wavelengths displaying Cooper
minima to avoid photoionization~\cite{AdamsPhoto, SaffmanFORT}.

In this paper we have calculated the radial matrix elements of arbitrary bound-bound,
bound-free and free-free transitions between \textit{S, P, D} and \textit{F}
states of alkali-metal atoms using the quasiclassical model by Dyachkov and
Pankratov (DP model)~\cite{Dyachkov1994, DyachkovFree}. In section II we present
examples of the Cooper minima for bound-free and bound-bound transitions and discuss the
accuracy of the theoretical method. Numerical
results are presented in Section III as density plots, revealing the Cooper
minima in bound-bound, bound-free and free-free transitions.

\section{The quasiclassical model and its applicability}

Radial matrix elements of the electric dipole transitions between arbitrary
atomic states (e.g., bound-bound or bound-free transitions) are required to
calculate the spectroscopic properties of atoms, including oscillator
strengths, lifetimes, photoionization cross-sections, and rates of
collisional ionization.

Although alkali-metal atoms have a single valence electron, only states with
small quantum defects exhibit truly hydrogen-like behavior. Due to
non-hydrogenic character of alkali-metal atoms, the calculation of radial
matrix elements remains a challenging task, since no exact analytical
solution for arbitrary transitions is available yet~\cite{BetheSalpeter}. The
oscillator strengths for alkali-metal atoms can strongly deviate from the
values for hydrogen. Accurate calculation of the radial integral for
transitions between states with small angular momentum are difficult because
of the need to take into account the interaction of the valence electron
with the atomic core.

A method based on the Coulomb approximation relies on the idea that the
Rydberg electron is localized mostly outside the atomic core, where the
potential is Coulombic. In the Numeric Coulomb Approximation~\cite{NCA} the
radial wavefunctions are obtained by solution of the Shrodinger equation
with the exact energies of the alkali-metal quantum states, expressed
through the quantum defect (Rydberg-Ritz formula, atomic units are used in this paper):

\begin{equation}
\label{eq1}
E_{n} = - \frac{{1}}{{2n_{eff}^{2}} }.
\end{equation}

%=========================================================================================================
%=========================================================================================================

\noindent Here $n_{eff} = n - \mu _{L} $ is an effective quantum number, $\mu _{L} $
is the quantum defect. Quantum defect accounts for the penetration of the
valence electron into the ionic core of a Rydberg atom. The quantum defects
are used as input parameter for the calculations and the integration is
truncated at the inner core radius.

Alternative forms of the Coulomb approximation were developed in~\cite{MCA}. The
Modified Coulomb Approximation (MCA) is a generalization of the analytical
expression for the hydrogen radial integral for non-integer quantum numbers.
It allows direct calculation of the radial matrix elements without numeric
integration.

Further simplification of the calculations in the Coulomb approximation is
achieved by extension of the quasiclassical approximation to the states with
low principal quantum numbers~\cite{DyachkovFree, Dyachkov1994}. The
radial matrix elements are expressed through transcendental functions, which
help to avoid inaccurateness of the direct numerical integration. This
method can substantially improve both the speed and reliability of the
calculations. However, the validity of most quasiclassical models was
restricted by transitions between neighboring excited states~\cite{Bureeva,
Naccache, Edmonds, DavydkinZon, GDK1994}.

In our previous works~\cite{BeterovLifetimes, BeterovNJP} we used the
quasiclassical model developed by Dyachkov and Pankratov~\cite{Dyachkov1994,DyachkovFree}. Their original approach provides more precise
values for the wave functions of the Rydberg and continuum states, compared
to the other quasiclassical models. Good agreement with numeric results
based on NCA model~\cite{NCA} or various model potential methods~\cite{
Ovsiannikov2, Marinescu} is observed.

Radial matrix elements for transitions between excited states of alkali-metal atoms display numerous features in their depenencies on $n$.
For example, minima were revealed in the transition probabilities for $nD \to {n}'F$ in
cesium~\cite{TheodosiouMin1980} and for $nD \to {n}'P$ in potassium. Hoogenraad
et~al.~\cite{Hoogenraad1995} observed theoretically and experimentally a Cooper
minimum for $nS \to {n}'P$ transitions in lithium.
To study the validity of the quasiclassical model for bound-bound transitions, we have calculated the transition probabilities for $nL \to {n}'{L}'$
transitions in alkali-metal atoms with $n < 100$ and $L,{L}' = 1,2,3$ using
the DP quasiclassical model, which is depicted in Appendix~B. Figure~\ref{BoundMin} shows the dependencies of the radial
matrix elements on the principal quantum number \textit{n} for $60F \to nD$
transitions in rubidium and cesium [Fig.~\ref{BoundMin}(a) and Fig.~\ref{BoundMin}(b), respectively], $60D
\to nP$ transitions in potassium [Fig.~\ref{BoundMin}(c)] and $60P \to nS$ transitions in
lithium [Fig.~\ref{BoundMin}(d)]. The observed minima for lithium, potassium and cesium are in
agreement with the results of Refs.~\cite{TheodosiouMin1980, HoogenraadDipole}.
The minima for $nF \to {n}'D$ transitions in rubidium appear only for high
\textit{n}, and have not been located yet, to the best of our knowledge.
These minima lie in the microwave region of about 150 GHz and could be
studied using microwave spectroscopy~\cite{Gallagher}.

%=========================================================================================================
%=========================================================================================================
%=========================================================================================================
\begin{figure}
\includegraphics[scale=0.38]{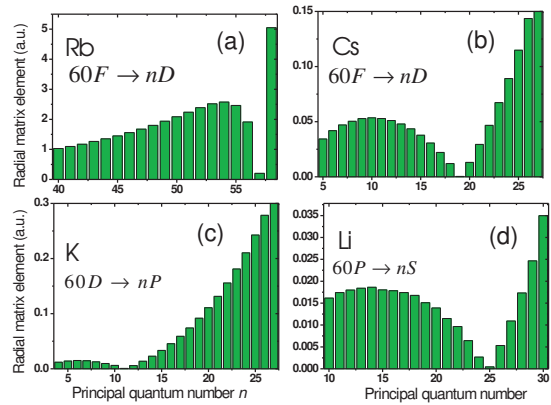}
\caption{
\label{BoundMin}(Color online).
The calculated radial matrix elements for bound-bound  (a) $60F \to nD$ transitions in rubidium; (b) $60F \to nD$ 
transitions in cesium; (c) $60D \to nP$ 
transitions in potassium; (d) $60P \to nS$ 
transitions in lithium.}
\end{figure}
%=========================================================================================================
%=========================================================================================================
%=========================================================================================================

Direct measurement of the radial matrix elements is of great importance for verification of the theory. However, due to lack of available experimental data for transitions between excited states of alkali-metal atoms, new measurements are required. In order to benchmark the model, we have earlier measured  the reduced matrix element for the diffuse series of rubidium~\cite{Piotrowicz2011}. 

We observed the Autler Townes splitting in a sample of ultra-cold Rb atoms using a 3-level ladder system. Briefly, we monitored the absorption of a weak probe laser scanned over the $5S_{1/2} \to 5P_{3/2}$ whilst simultaneously a strong coupling laser, locked to the $ 5P_{3/2}\to nD_{5/2}$ transition illuminated the atoms. The strong coupling laser gave rise to two absorption peaks separated by the Rabi frequency of the atom-coupling laser interaction. Knowledge of the laser intensity allowed us to measure the dipole matrix elements of the $5P_{3/2}\to nD_{5/2}$ transitions to within $7\%$ accuracy.

\begin{figure}
\includegraphics[scale=0.4]{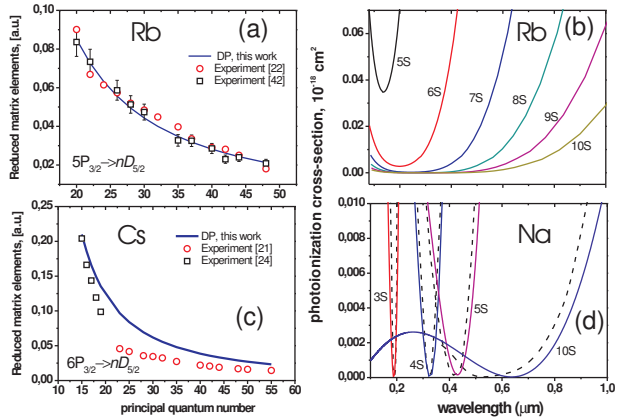}
\caption{
\label{CooperPhoto}(Color online).
(a), (c) Comparison of the calculated reduced dipole moments for (a) rubidium atoms with 
experiment~\cite{Nadeem2011, Piotrowicz2011} and (c) cesium atoms with experiment~\cite{Haq2010, Fabry}.
(b), (d) Cooper minima in photoionization cross-sections of $nS$ (b)~rubidium and (d)~sodium atoms. Solid
curves - this work. Broken curves - theoretical calculations from
Ref.~\cite{Aymar1978}; 
}
\end{figure}

We have also compared the reduced dipole moments calculated using DP model with
other available  experimental data on diffuse series of rubidium~\cite{Nadeem2011,Piotrowicz2011} and cesium~\cite{Haq2010,Fabry}. Good agreement with experiments of Refs.~\cite{Piotrowicz2011, Nadeem2011} is confirmed in Fig.~\ref{CooperPhoto}(a) for rubidium $5P
\to nD$ transitions. For cesium the theoretical values in Fig.~\ref{CooperPhoto}(c) are in
agreement with the experiment only for the lowest \textit{nD} states and
differ from the experimental values for higher $n$ by a factor of two. Our
theoretical results for cesium are, however, in excellent agreement with the
previous calculations of Ref.~\cite{Fabry}. The observed disagreement between experiment and theory can result from improper
account for core polarization for heavy cesium atoms~\cite{Migdalek}. However, we
expect that for transitions between the states with larger principal quantum
numbers the accuracy of the semiclassical approximation will be
significantly increased due to smaller interaction of the Rydberg electron with the atomic core.
%=========================================================================================================
%=========================================================================================================
%=========================================================================================================
%=========================================================================================================
%=========================================================================================================
%=========================================================================================================

%=========================================================================================================
%=========================================================================================================
%=========================================================================================================
\begin{figure}
\includegraphics[scale=0.5]{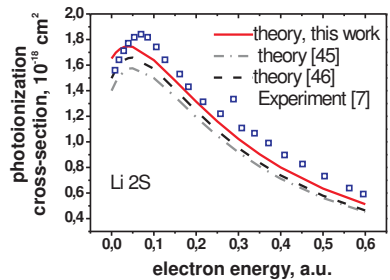}
\caption{
\label{Lithium}(Color online).
Lithium 2S photoionization cross-section. Comparison of the quasiclassical calculations with the 
experiment~\cite{HudsonCarter1967} and theory~\cite{LiNaPlasmaSahoo, Peach}.}
\end{figure}
%=========================================================================================================
%=========================================================================================================
%=========================================================================================================

Calculated Cooper minima in the photoionization cross-sections of the \textit{nS}
states of rubidium and sodium are shown in Fig.~\ref{CooperPhoto}(b) and (d), respectively.
The solid curves in Fig.~\ref{CooperPhoto}(b) and (d) represent the photoionization cross-sections
calculated using our model based on the DP method. In the case of sodium,
these are compared with the quantum-mechanical calculations of
Ref.~\cite{Aymar1978}, shown as broken curves in Fig.~\ref{CooperPhoto}(d). Good agreement is
observed for sodium \textit{nS} states with $n < 5$. For higher states
the positions of the Cooper minimum are significantly shifted. In Ref.~\cite{Bezuglov1999} it has been argued that the discrepancy in the quasiclassical
calculations could be corrected by adjusting the phase in the radial
integral in order to compensate the phase shift of the radial wavefunction
from the value given by the quantum defect. At the same time, the
reliability of the quasiclassical approximation is also expected to improve
with the increase of the principal quantum number, and only experimental
data could confirm the validity of the theory.

In figure~\ref{Lithium} we have compared the calculated photoionization cross-section
for lithium 2S state with the experiment of Ref.~\cite{HudsonCarter1967} and theory
of Refs.~\cite{LiNaPlasmaSahoo, Peach}. It is seen that our approach provides a better agreement with the experiment.

We conclude that the DP model is suitable for calculation of the radial
matrix elements and photoionization cross-sections with an accuracy better
than a factor of two for low \textit{n} and much enhanced for higher excited
states, as confirmed by the good agreement between experiment and theory in
recent lifetime measurements~\cite{Tate, Feng}.

\section{Maps of the cooper minima}

Using the quasiclassical model of Delone et al.~\cite{GDK1994}, in Ref.~\cite{HoogenraadDipole} it has been shown that the radial matrix elements for
bound-bound, bound-free and free-free transitions could be expressed in a
universal way through the numerically calculated relative matrix elements
$R_{rel} $, multiplied by the appropriate normalization factors:

\begin{align}
\label{eq2}
&R\left( {nL \to {n}'{L}'} \right) = \frac{{0.4108 \times R_{rel} \left(
{E_{n} L \to E_{n'}{L}'} \right)}}{{n_{eff}^{'3/2} \times n_{eff}^{3/2} \times
\left| {E_{n'} - E_{n}}  \right|^{5/3}}} = &\nonumber\\
 &= \frac{{0.4108 \times R_{rel} \left( {E_{n} L \to E_{n'}{L}'}
\right)}}{{\left( { - 2E_{n'}}  \right)^{- 3/4} \times \left( {- 2E_{n}}
\right)^{ - 3/4} \times \left| {E_{n'}- E_{n}}  \right|^{5/3}}} &\nonumber\\
 &R\left( {nL \to {E}'{L}'} \right) = \frac{{0.4108 \times R_{rel} \left(
{E_{n} L \to {E}'{L}'} \right)}}{{n_{eff}^{3/2} \times \left| {{E}' - E_{n}
} \right|^{5/3}}} =&\nonumber \\
 &= \frac{{0.4108 \times R_{rel} \left( {E_{n} L \to {E}'{L}'}
\right)}}{{\left( { - 2E_{n}}  \right)^{ - 3/4} \times \left| {{E}' - E_{n}
} \right|^{5/3}}} &\nonumber\\
&R\left( {EL \to {E}'{L}'} \right) = \frac{{0.4108 \times R_{rel} \left( {EL
\to E'L'} \right)}}{{\left| {{E}' - E} \right|^{5/3}}} &
 \end{align}

\noindent The prefactor $\left( {4/3} \right)^{1/3}/\Gamma \left( {1/3} \right) =
0.4108$ results from the asymptotic expression for the quasiclassical matrix
elements for $n \to n + 1$ transitions \cite{HoogenraadDipole}. Relative matrix
elements $R_{rel} \left( {EL \to {E}'{L}'} \right)$ introduced in Eq.(\ref{eq2}) are
convenient as these are slowly varying functions of $E$ and ${E}'$. The
dependence of $R_{rel} \left( {EL \to {E}'{L}'} \right)$ on the energy
${E}'$ of the final state passes smoothly through the ionization threshold
\cite{HoogenraadDipole}. The asymptotic $\left| {{E}' - E} \right|^{ - 5/3}$
dependence of the radial matrix elements is incorrect for transitions
between neighboring states with $E \approx {E}'$, where the dipole matrix
elements rapidly increase~\cite{HoogenraadDipole}. In this case the radial matrix
elements can be calculated numerically using a DP model~\cite{Dyachkov1994,
DyachkovFree}, or NCA~\cite{NCA}.

We have calculated the relative matrix elements $R_{rel} \left( {E_{n} L \to
E_{{n}'} {L}'} \right)$, $R_{rel} \left( {E_{n} L \to {E}'{L}'} \right)$,
$R_{rel} \left( {EL \to {E}'L} \right)$ for transitions between \textit{S},
\textit{P}, \textit{D} and \textit{F} states of alkali-metal atoms, starting
from the ground state. The energies of the continuum states were taken
within $0 < E < 0.5$ (atomic units) to extend the results of Ref.
\cite{HoogenraadDipole} to the area where Cooper minima are expected for
alkali-metal atoms.

%=========================================================================================================
%=========================================================================================================
%=========================================================================================================
\begin{figure}
\includegraphics[scale=0.8]{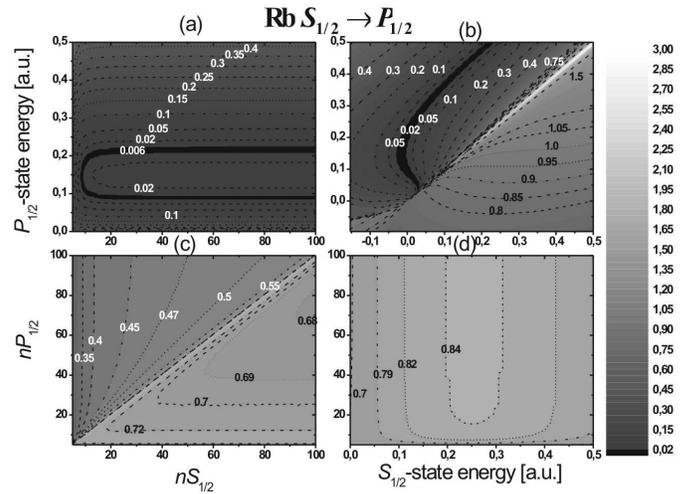}
\caption{
\label{RbSP}(Color online).
Density plots of the relative matrix elements for (a) Rb bound-free $nS_{1/2} \to
{E}'P_{1/2} $ transitions; (b) arbitrary Rb $ES_{1/2} \to {E}'P_{1/2}$
transitions including discrete and continuum spectra; (c) Rb bound-bound
$nS_{1/2} \to {n}'P_{1/2}$  transitions; (d) Rb bound-free
$nP_{1/2} \to {E}'S_{1/2}$ transitions.}
\end{figure}
%=========================================================================================================
%=========================================================================================================
%=========================================================================================================

%=========================================================================================================
%=========================================================================================================
%=========================================================================================================
\begin{figure}
\includegraphics[scale=0.8]{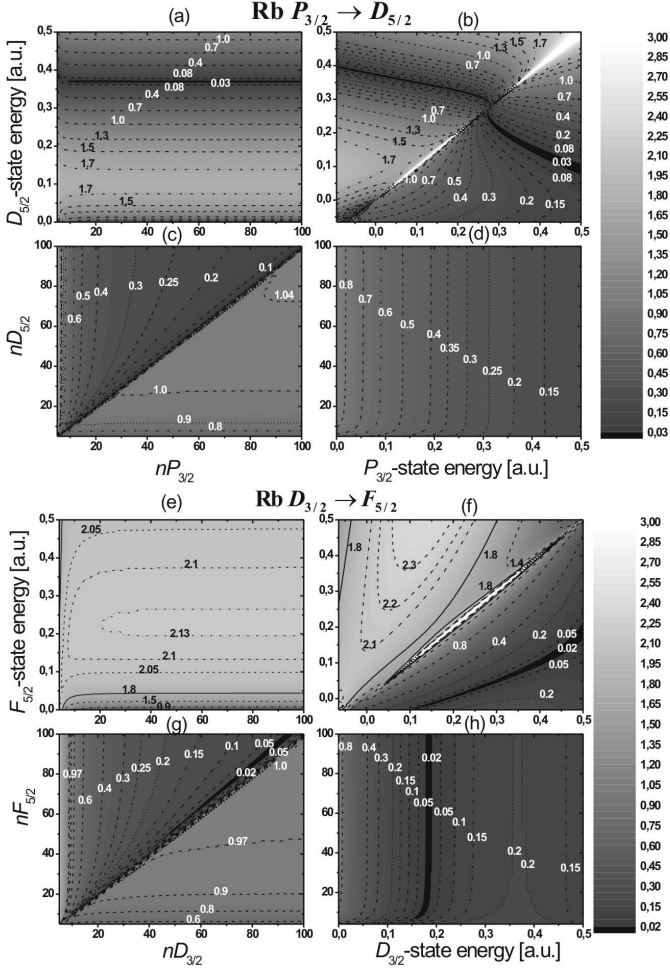}
\caption{
\label{RbPDF}(Color online).
Density plots of the relative matrix elements in Rb atoms for (a) bound-free $nP_{3/2} \to
{E}'D_{5/2}$ transitions; (b) all $EP_{3/2} \to {E}'D_{5/2}$ transitions;
(c) bound-bound $nP_{3/2} \to {n}'D_{5/2}$ transitions; (d) bound-free
$nD_{5/2} \to {E}'P_{3/2}$ transitions; (e) bound-free $nD_{3/2} \to
{E}'F_{5/2}$ transitions; (f) all $ED_{3/2} \to {E}'F_{5/2}$ transitions;
(g) bound-bound $nD_{3/2} \to {n}'F_{5/2}$ transitions; (h) bound-free
$nF_{5/2} \to {E}'D_{3/2}$ transitions. }
\end{figure}
%=========================================================================================================
%=========================================================================================================
%=========================================================================================================
%=========================================================================================================
%=========================================================================================================
\begin{figure}
\includegraphics[scale=0.8]{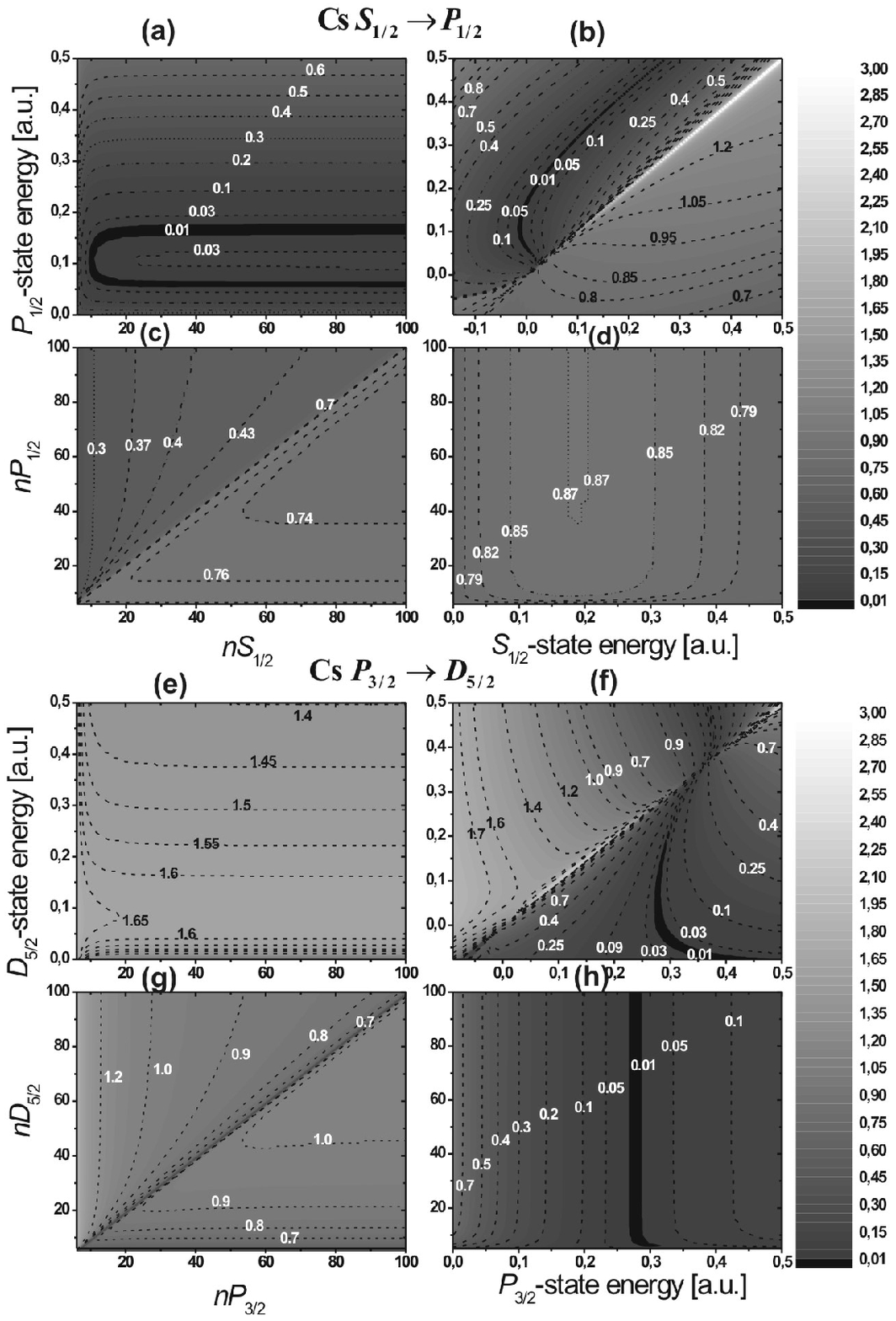}
\caption{
\label{CsSPD}(Color online).
Density plots of the relative matrix elements in Cs atoms for (a) bound-free
$nS_{1/2} \to {E}'P_{1/2} $ transitions; (b) all $ES_{1/2} \to {E}'P_{1/2} $
transitions; (c) bound-bound $nS_{1/2} \to {n}'P_{1/2}$ transitions; (d)
bound-free $nP_{1/2} \to {E}'S_{1/2}$ transitions; (e) bound-free $nP_{3/2}
\to {E}'D_{5/2}$ transitions; (f) all $EP_{3/2} \to {E}'D_{5/2} $
transitions; (g) bound-bound $nP_{3/2} \to {n}'D_{5/2}$ transitions; (h)
bound-free $nD_{5/2} \to {E}'P_{3/2}$ transitions.}
\end{figure}
%=========================================================================================================
%=========================================================================================================
%=========================================================================================================

%=========================================================================================================
%=========================================================================================================
%=========================================================================================================
\begin{figure}
\includegraphics[scale=0.8]{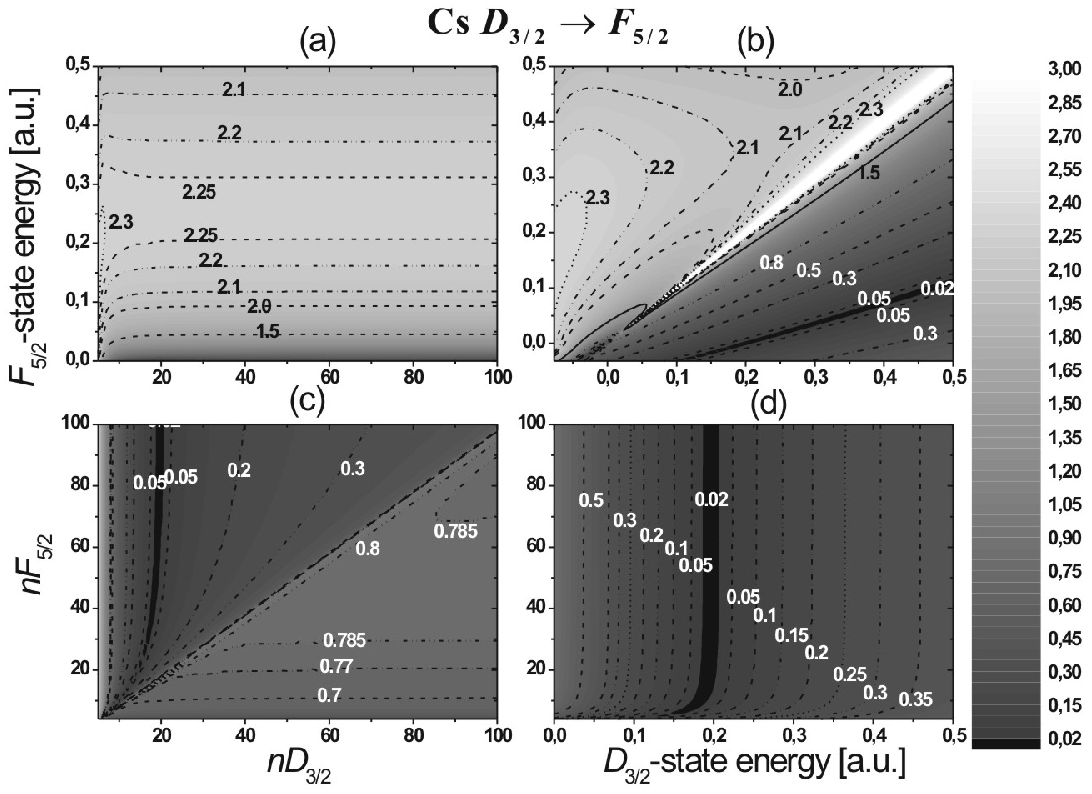}
\caption{
\label{CsDF}(Color online).
Density plots of the relative matrix elements in Cs atoms for (a) bound-free $nD_{3/2} \to
{E}'F_{5/2} $ transitions; (b) all $ED_{3/2} \to {E}'F_{5/2} $ transitions;
(c) bound-bound $nD_{3/2} \to {n}'F_{5/2}$ transitions; (d) bound-free
$nF_{5/2} \to {E}'D_{3/2}$ transitions.}
\end{figure}
%=========================================================================================================
%=========================================================================================================
%=========================================================================================================

%=========================================================================================================
%=========================================================================================================
\begin{figure}
\includegraphics[scale=0.8]{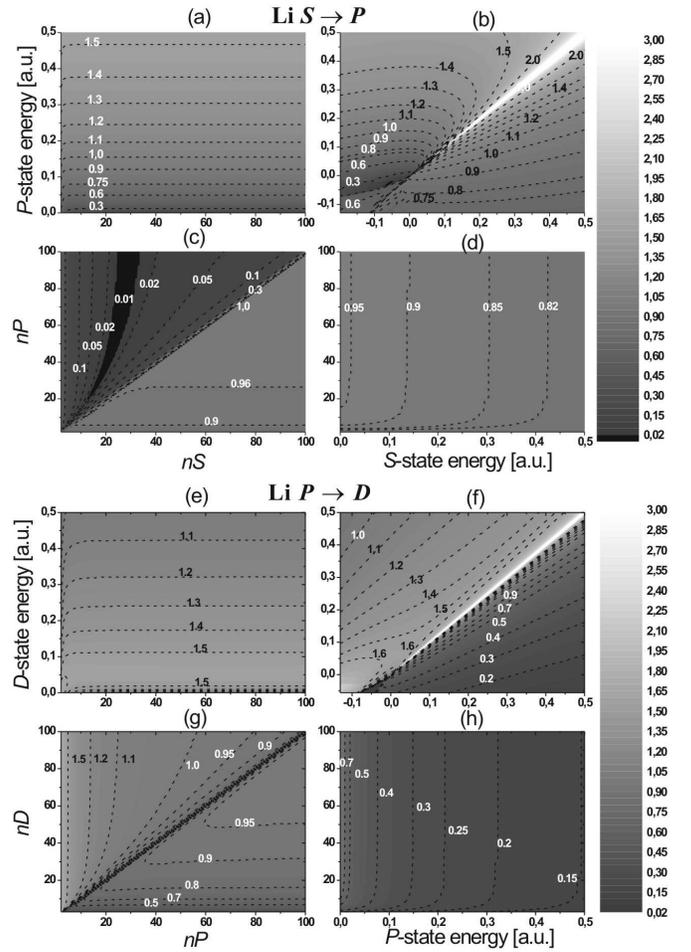}
\caption{
\label{LiSPD}(Color online).
Density plots of the relative matrix elements in Li atoms for (a) bound-free $nS \to {E}'P$
transitions; (b) all $ES \to {E}'P$ transitions; (c) bound-bound $nS \to
{n}'P$ transitions; (d) bound-free $nP \to {E}'S$ transitions; (e) bound-free
$nP \to {E}'D$ transitions; (f) all $EP \to {E}'D$ transitions; (g)
bound-bound $nP \to {n}'D$ transitions; (h) bound-free $nD \to {E}'P$
transitions.}
\end{figure}
%=========================================================================================================
%=========================================================================================================
%=========================================================================================================

%=========================================================================================================
%=========================================================================================================
\begin{figure}
\includegraphics[scale=0.8]{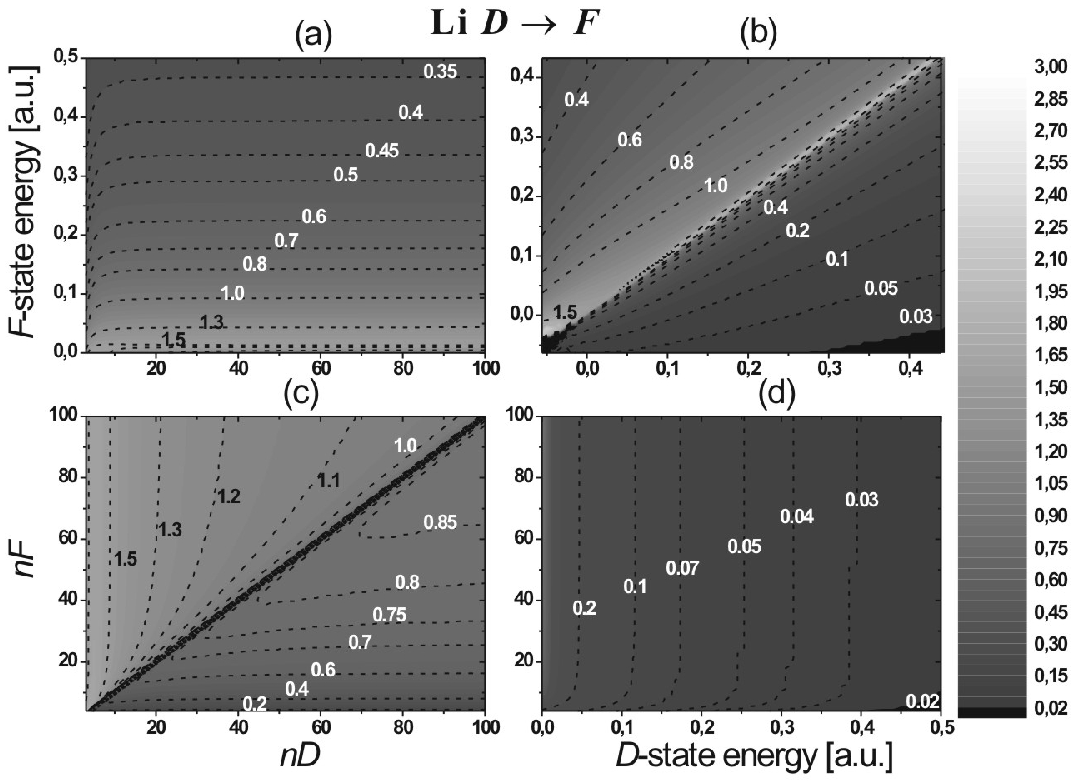}
\caption{
\label{LiDF}(Color online).
Density plots of the relative matrix elements in Li atoms for (a) bound-free $nD \to
{E}'F$ transitions; (b) all $ED \to {E}'F$ transitions; (c) bound-bound $nD
\to {n}'F$ transitions; (d) bound-free $nF \to {E}'D$ transitions.}
\end{figure}
%=========================================================================================================
%=========================================================================================================
%=========================================================================================================

%=========================================================================================================
%=========================================================================================================
\begin{figure}
\includegraphics[scale=0.8]{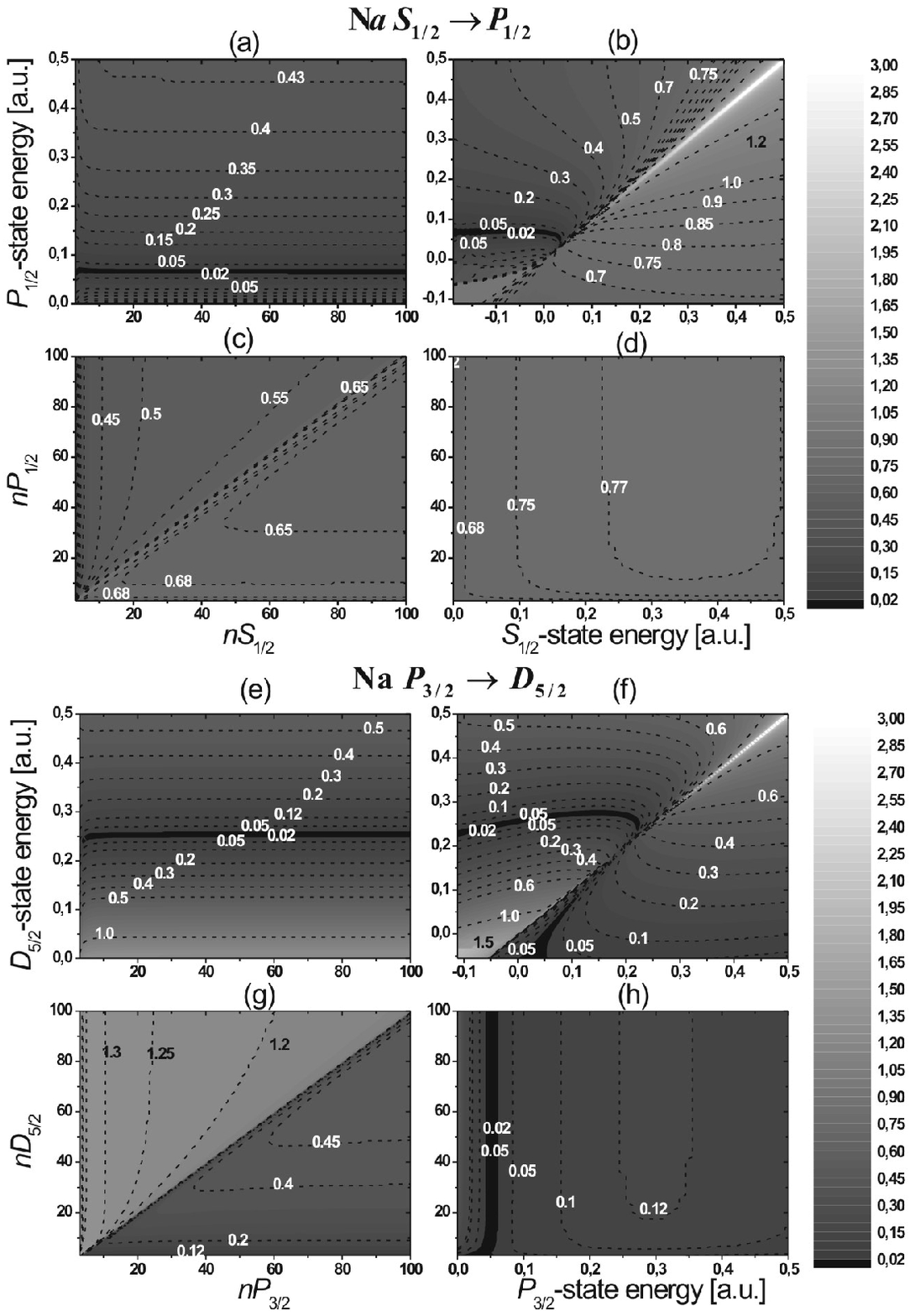}
\caption{
\label{NaSPD}(Color online).
Density plots of the relative matrix elements in Na atoms for (a) bound-free $nS_{1/2} \to
{E}'P_{1/2}$ transitions; (b) all $ES_{1/2} \to {E}'P_{1/2}$ transitions;
(c) bound-bound $nS_{1/2} \to {n}'P_{1/2}$ transitions; (d) bound-free
$nP_{1/2} \to {E}'S_{1/2}$ transitions; (e) bound-free $nP_{3/2} \to
{E}'D_{5/2}$ transitions; (f) all $EP_{3/2} \to {E}'D_{5/2}$ transitions;
(g) bound-bound $nP_{3/2} \to {n}'D_{5/2}$ transitions; (h) bound-free
$nD_{5/2} \to {E}'P_{3/2} $ transitions.}
\end{figure}
%=========================================================================================================
%=========================================================================================================
%=========================================================================================================

%=========================================================================================================
%=========================================================================================================
\begin{figure}
\includegraphics[scale=0.8]{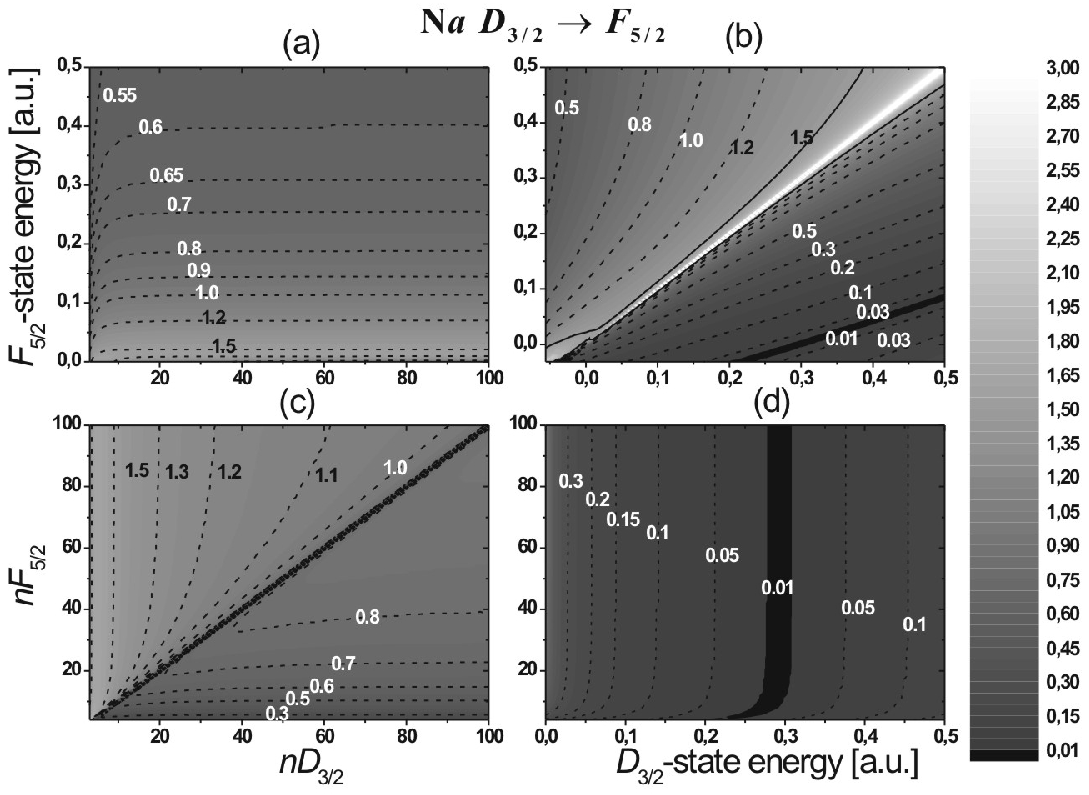}
\caption{
\label{NaDF}(Color online).
Density plots of the relative matrix elements in Na atoms for (a) bound-free $nD_{3/2} \to
{E}'F_{5/2} $ transitions; (b) all $ED_{3/2} \to {E}'F_{5/2} $ transitions;
(c) bound-bound $nD_{3/2} \to {n}'F_{5/2}$ transitions; (d) bound-free
$nF_{5/2} \to {E}'D_{3/2} $ transitions;}
\end{figure}
%=========================================================================================================
%=========================================================================================================
%=========================================================================================================

%=========================================================================================================
%=========================================================================================================
\begin{figure}
\includegraphics[scale=0.8]{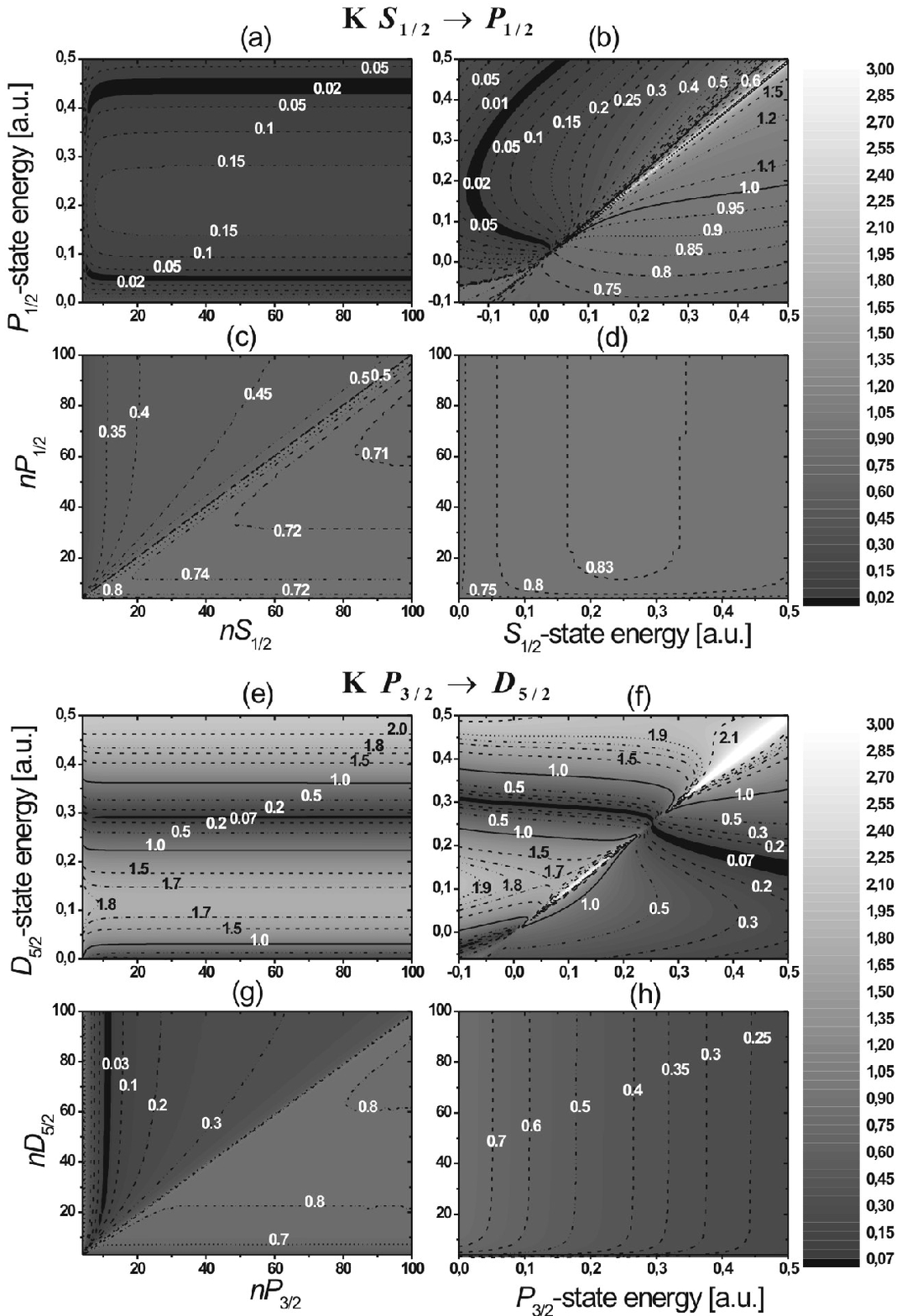}
\caption{
\label{KSPD}(Color online).
Density plots of the relative matrix elements in K atoms for (a) bound-free $nS_{1/2} \to
{E}'P_{1/2} $ transitions; (b) all $ES_{1/2} \to {E}'P_{1/2} $ transitions;
(c) bound-bound $nS_{1/2} \to {n}'P_{1/2}$ transitions; (d) bound-free
$nP_{1/2} \to {E}'S_{1/2} $ transitions; (e) bound-free $nP_{3/2} \to
{E}'D_{5/2} $ transitions; (f) all $EP_{3/2} \to {E}'D_{5/2}$ transitions;
(g) bound-bound $nP_{3/2} \to {n}'D_{5/2}$ transitions; (h) bound-free
$nD_{5/2} \to {E}'P_{3/2}$ transitions.}
\end{figure}
%=========================================================================================================
%=========================================================================================================
%=========================================================================================================

%=========================================================================================================
%=========================================================================================================
\begin{figure}
\includegraphics[scale=0.8]{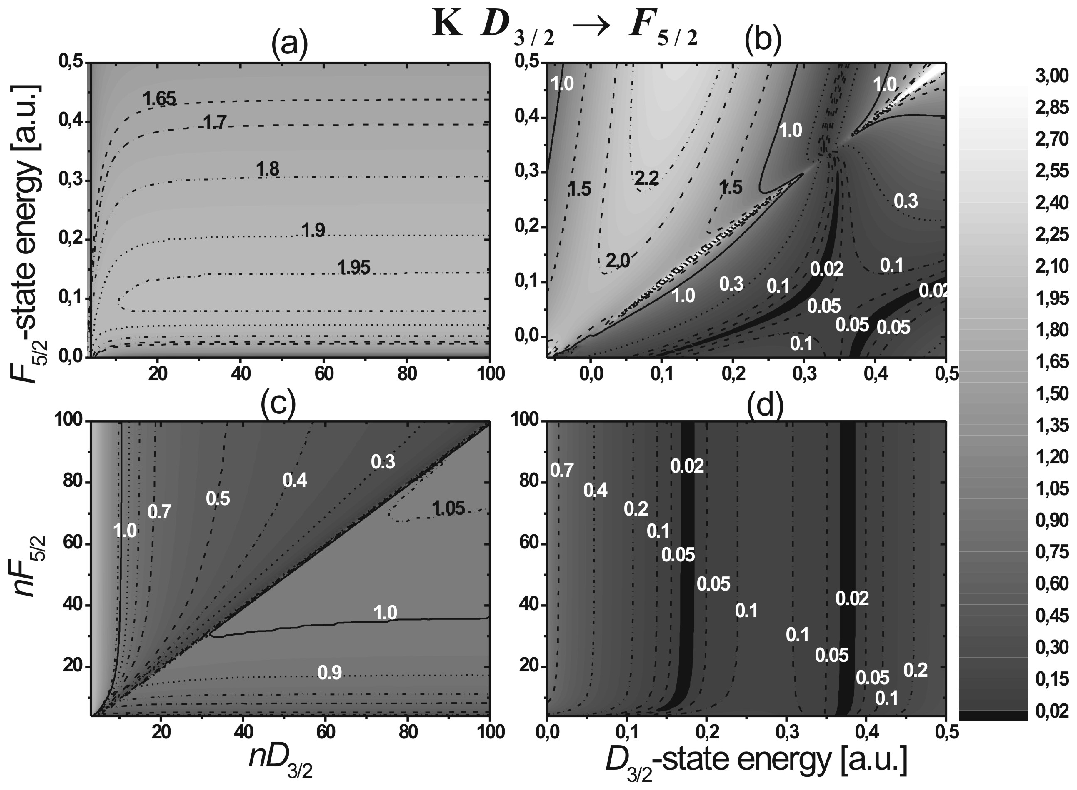}
\caption{
\label{KDF}(Color online).
Density plots of the relative matrix elements in K atoms for (a) bound-free
$nD_{3/2} \to {E}'F_{5/2} $ transitions; (b) all $ED_{3/2} \to {E}'F_{5/2} $
transitions; (c) bound-bound $nD_{3/2} \to {n}'F_{5/2}$ transitions; (d)
bound-free $nF_{5/2} \to {E}'D_{3/2}$ transitions.}
\end{figure}
%=========================================================================================================
%=========================================================================================================
%=========================================================================================================

\subsection{Rubidium}

Rubidium and cesium atoms are widely employed  in laser cooling experiments, and
we shall discuss them in more detail. The relative matrix elements $R_{rel}
\left( {S_{1/2} \to P_{1/2}}  \right)$ for rubidium are shown in Fig.~\ref{RbSP}.
Following  Ref.~\cite{HoogenraadDipole}, we present our numerical
results as density plots. We use both \textit{E}-scaled and
\textit{n}-scaled plots, since the latter are more appropriate to the
relative matrix elements for bound-bound and bound-free transitions from
states with large principal quantum numbers \textit{n.} The signs of
the radial matrix elements are unimportant in the calculation of transition
probabilities and photoionization cross-sections, therefore we present only their
absolute values.

Figure~\ref{RbSP}(a) shows the relative matrix elements $R_{rel} \left( {nS_{1/2} \to
{E}'P_{1/2}}  \right)$ for bound-free transitions in rubidium atoms. The
horizontal axis is the principal quantum number of the \textit{nS} states,
while the vertical axis is the binding energy ${E}'$ of the continuum
\textit{P} states (in atomic units). From Fig.~\ref{RbSP}(a) one finds that for a
bound-free transition between the Rydberg 40\textit{S}$_{1/2}$ state and the
continuum ${E}'P_{1/2} $ state with ${E}' = 0.18$ the relative matrix
element is $R_{rel} \left( {nS_{1/2} \to {E}'P_{1/2}}  \right) = 0.02$.
According to Eq.(\ref{eq1}) and Table~1 in the appendix~A the energy of the 40S$_{1/2
}$ state is $E_{n} = - 3.68 \times 10^{-4}$, and the energy difference is
$\left| {{E}' - E_{n}}  \right| = 0.18$. Then the absolute radial matrix
element can be found from Eq.(\ref{eq2}): $R\left( {40S_{1/2} \to {E}'P_{1/2}}
\right) = 6.1 \times 10^{ - 4}$. To highlight Cooper minima, the regions where the relative matrix
element $R_{rel} \left( {nS_{1/2} \to {E}'P_{1/2}}  \right)$ falls below
0.02 are filled by black. The interesting feature of Fig.~\ref{RbSP}(a) is the
presence of the two sharp Cooper minima at ${E}' = 0.09$ and ${E}' = 0.21$
for $n >10$.

Figure~\ref{RbSP}(b) shows the relative matrix elements for all possible bound-bound,
bound-free and free-free $R_{rel} \left( {S_{1/2} \to P_{1/2}}  \right)$
transitions in rubidium atoms, plotted in the energy scale. The horizontal axis carries the binding
energy of \textit{S} states while the
binding energy of \textit{P} states is given in   the vertical axis.
For a bound-free transition between the
6\textit{P}$_{1/2}$ state with $E = - 0.045$ and the continuum
\textit{S}$_{1/2}$ state with ${E}' = 0.15$, the relative matrix element is
$R_{rel} \left( {E_{n} P_{1/2} \to {E}'S_{1/2}}  \right) = 0.8$. Then Eq.~(\ref{eq2})
gives the absolute radial matrix element $R\left( {E_{n} P_{1/2} \to
{E}'S_{1/2}}  \right) = 0.82$ (the energy difference is $\left| {{E}' - E}
\right| = 0.195$). The same procedure can be applied to calculate radial
matrix elements for all bound-bound, bound-free and free-free transitions.

A prominent Cooper minimum is observed in Fig.~\ref{RbSP}(a),~(b) for the bound-free
$nS_{1/2} \to {E}'P_{1/2}$ and free-free $ES_{1/2} \to {E}'P_{1/2} $
transitions in rubidium. For the \textit{nS}$_{1/2}$ states with $E_{n} > -
0.03$ (corresponding to $n>7$) the relative matrix elements fall down
below 0.02, while for the lower \textit{nS}$_{1/2}$ states the minimum is
not so sharp.

The relative matrix elements $R_{rel} \left( {E_{n} P_{1/2} \to E_{{n}'} S_{1/2}
} \right)$ for bound-bound transitions in rubidium  are presented in
Fig.~\ref{RbSP}(c).  Since the
relative matrix elements slowly vary with $n$ and ${n}'$, accurate
calculation of the radial matrix elements from the data of Fig.~\ref{RbSP}(c) is
possible as described earlier. For example, $R_{rel} \left( {27S_{1/2} \to
80P_{1/2}}  \right) = 0.45$, $E_{27S} = - 8.776 \times 10^{ - 4}$, $E_{80P}
= - 8.36 \times 10^{ - 5}$ and the energy difference is $\left| {E_{{n}'} -
E_{n}}  \right| = 7.94 \times 10^{ - 4}$. From Eq.~(\ref{eq2}) one finds $R\left(
{27S_{1/2} \to 80P_{1/2}}  \right) = 0.342$.

Relative matrix elements $R_{rel} \left( {nP_{1/2} \to {E}'S_{1/2}}
\right)$ for bound-free transitions in rubidium atoms are shown in Fig.~\ref{RbSP}(d).
 As an example, the relative
matrix element is $R_{rel} \left( {60P_{1/2} \to {E}'S_{1/2}}  \right) =
0.84$ for ${E}' = 0.2$; then the radial matrix element is $R\left(
{60P_{1/2} \to {E}'S_{1/2}}  \right) = 0.012$.

Figure~\ref{RbPDF} displays the relative matrix elements $R_{rel} \left( {P_{3/2} \to
D_{5/2}}  \right)$ and $R_{rel} \left( {D_{3/2} \to F_{5/2}}  \right)$ for
rubidium atoms in the same way as in Fig.~\ref{RbSP}. Relative matrix elements for
other fine-structure components of the rubidium \textit{P} and \textit{D}
states are not presented, since the difference between them is too small to
be distinguishable on our density plots. Cooper minima are observed for
bound-free $nP_{3/2} \to {E}'D_{5/2} $ transitions with ${E}' \approx 0.37$
[Fig.~\ref{RbPDF}(a)], free-free $EP_{3/2} \to {E}'D_{5/2} $ and $ED_{3/2} \to F_{5/2}$ transitions [Fig.~\ref{RbPDF}(b)],
and bound-free $nF_{5/2} \to {E}'D_{3/2} $ transitions with ${E}' \approx
0.17$ [Fig.~\ref{RbPDF}(h)].

%=========================================================================================================

\subsection{Cesium}

Figure~\ref{CsSPD} shows relative matrix elements $R_{rel} \left( {S_{1/2} \to P_{3/2}
} \right)$ and $R_{rel} \left( {P_{3/2} \to D_{5/2}}  \right)$ for cesium
atoms. Relative matrix elements $R_{rel} \left( {D_{3/2} \to F_{5/2}}
\right)$ for cesium atoms are presented in Fig.~\ref{CsDF}. Two Cooper minima are
observed for bound-free $nS_{1/2} \to {E}'P_{1/2} $ transitions with ${E}'
\approx 0.06$ and ${E}' \approx 0.17$ in Fig.~\ref{CsSPD}(a). The Cooper minima are also
noticed for the bound-free $n D_{5/2}\to EP_{3/2} $ transitions with ${E}'
\approx 0.28$ [Fig.~\ref{CsSPD}(h)], bound-free $nF_{5/2} \to {E}'D_{3/2} $ transitions
with ${E}' = 0.19$[Fig.~\ref{CsDF}(h)], and bound-bound $nD_{3/2} \to {n}'F_{5/2}
$ transitions with $n<23$ [Fig.~\ref{CsDF}(g)].

Cooper minimum in the discrete spectrum was first discussed in Ref.~\cite{TheodosiouMin1980}. In Ref.~\cite{HoogenraadDipole}  a
continuation of this minimum in the bound-free $nD_{3/2} \to {E}'F_{5/2} $
transitions was found,  with the energy ${E}' \approx 10^{ - 3}$ being close to the
ionization threshold. These features were reproduced in our calculations,
but they are not shown due to the large energy scale of our density plots.
The near-threshold Cooper minima for Rydberg states of alkali-metal atoms
were also discussed in Ref.~\cite{HoogenraadDipole}.

\subsection{Lithium}

For lithium atoms the relative matrix elements obtained for $R_{rel} \left(
{S \to P} \right)$ and $R_{rel} \left( {P \to D} \right)$ transitions are
shown in Fig.~\ref{LiSPD}, and for $R_{rel} \left( {D \to F} \right)$ transitions in
Fig.~\ref{LiDF}. Fine structure is neglected due to the small fine splitting. One may see
that the matrix elements of the bound-free $S \to P$ transitions slowly decrease
as the energy of the continuum state grows. A Cooper minimum is observed for
the bound-bound $nS_{1/2} \to {n}'P_{1/2} $ transitions [Fig.~\ref{LiSPD}(a)]. This
minimum has been found earlier in Ref.~\cite{Hoogenraad1995} in an experimental
study of the far-infrared transitions between Rydberg states. Later on we
have shown that such minimum can also appear in the BBR-induced transitions
\cite{BeterovJETP2008}.

\subsection{Sodium}

For sodium atoms the relative matrix elements obtained for $R_{rel} \left(
{S_{1/2} \to P_{1/2}}  \right)$ and $R_{rel} \left( {P_{3/2} \to D_{5/2}}
\right)$ transitions are shown in Fig.~\ref{NaSPD}, and for $R_{rel} \left( {D_{3/2}
\to F_{5/2}}  \right)$ transitions in Fig.~\ref{NaDF}. A Cooper minimum is observed
for the bound-free $nS_{1/2} \to {E}'P_{1/2} $ transitions at ${E}' \approx
0.06$ [Fig.~\ref{NaSPD}(a)], bound-free $nD_{5/2} \to {E}'P_{3/2} $ transitions at
${E}' \approx 0.05$ [Fig.~\ref{NaSPD}(h)], bound-free $nP_{3/2} \to {E}'D_{5/2} $
transitions at ${E}' \approx 0.25$ [Fig.~\ref{NaSPD}(e)], and bound-free $nF_{5/2} \to
{E}'D_{3/2} $ transitions at ${E}' \approx 0.29$ [Fig.~\ref{NaDF}(d)]. For $nS_{1/2}
\to {E}'P_{1/2} $ and $nD_{5/2} \to {E}'P_{3/2} $ transitions in sodium a
Cooper minimum is found to be close to the ionization threshold.

\subsection{Potassium}

For potassium atoms the relative matrix elements obtained for $R_{rel}
\left( {S_{1/2} \to P_{1/2}}  \right)$ and $R_{rel} \left( {P_{3/2} \to
D_{5/2}}  \right)$ transitions are shown in Fig.~\ref{KSPD}, and for $R_{rel} \left(
{D_{3/2} \to F_{5/2}}  \right)$ transitions in Fig.~\ref{KDF}. Interesting features
are observed in the radial matrix elements of bound-bound and bound-free
transitions. The two minima have been located in the relative matrix elements
of the bound-free $nS_{1/2} \to {E}'P_{1/2} $ transitions at ${E}' \approx
0.05$ and ${E}' \approx 0.45$ [Fig.~\ref{KSPD}(a)]. A Cooper minimum is also observed
for $nP_{3/2} \to {E}'D_{5/2} $ transitions at ${E}' \approx 0.29$ [Fig.~\ref{KSPD}(g)].
 A minimum in the discrete spectrum has been found [Fig.~\ref{KSPD}(c)],
which is similar to the lithium $S \to P$ and cesium $D \to F$ transitions.
This minimum in potassium atoms was first discussed by Theodosiou
\cite{TheodosiouMin1980}. Finally, for the bound-free $nF_{5/2} \to {E}'D_{3/2} $
transitions in potassium, a Cooper minimum is registered at ${E}' \approx
0.18$ and ${E}' \approx 0.38$ [Fig.~\ref{KDF}(d)].

\section{Conclusion}

The quasiclassical model developed by Dyachkov and Pankratov~\cite{Dyachkov1994,
DyachkovFree} can be used for fast and reliable calculations of the radial
matrix elements for bound-bound, bound-free and free-free transitions
between arbitrary states of alkali-metal atoms. We have demonstrated this by
perfoming the numerical calculations of the radial matrix elements for
transitions between \textit{S}, \textit{P}, \textit{D} and \textit{F} states
with the energies $E < 0.5$ (in atomic units) in all alkali-metal atoms. Our
results on radial matrix elements are in good agreement with  numerical calculations in the Coulomb
approximations~\cite{HoogenraadDipole}. Our theoretical results~\cite{BeterovLifetimes} are also
consistent with the experimental measurements of the effective lifetimes of
Rydberg states~\cite{Tate, Feng}, oscillator strengths~\cite{Piotrowicz2011,
Nadeem2011}, and photoionization cross-sections~\cite{HudsonCarter1967}. Our
approach allowed us to reveal several unknown Cooper minima, both in the discrete
and continuum spectra, which would be interesting to confirm experimentally. Reliability of the 
quasiclassical model for study of the Cooper minima is verified by good agreement with calculations of Ref.~\cite{TheodosiouMin1980} for bound-bound transitions
and satisfactory agreement with calculations of Ref.~\cite{Aymar1978} for bound-free transitions.
We conclude that the quasiclassical model of Dyachkov and Pankratov is a
universal method for systematic calculation of the radial matrix elements for transitions
between excited states of alkali-metal atoms.

\section{Acknowledgments}

This work was supported by Grants of the President of Russia MK.7060.2012.2, MK.3727.2011.2, RFBR Grant No.~10-02-00133, Russian Academy of Sciences
and the Dynasty foundation. SB , CM, AK and CMC aknowledge support from
EPSRC grant No.~EP/F031130/1.

\appendix
\section{The quantum defects.}

The values of the quantum defect can be obtained by fitting the
experimentally measured energies to Eq.(\ref{eq1}) \cite{Def1983}:

\begin{equation}
\label{eq3}
\mu _{L} \left({n} \right) = {a}'_{L} + \frac{{{b}'_{L}} }{{\mathord{n_{eff}^{2}}}}
+ \frac{{{c}'_{L}} }{{n_{eff}^{4}} } + \frac{{{d}'_{L}} }{{n_{eff}^{6}} } +
\frac{{{e}'_{L}} }{{n_{eff}^{8}} }...,
\end{equation}

\noindent where ${a}'_{L} $, ${b}'_{L} $, ${c}'_{L} $, ${d}'_{L} $, ${e}'_{L} $ are
the Rydberg-Ritz fitting coefficients. The quantum defects can also be
expressed through the modified Rydberg-Ritz coefficients, which were
tabulated for alkali-metal Rydberg atoms in Ref.~\cite{Def1983}:

\begin{align}
\label{eq4}
 &\mu _{L} \left( {n} \right) = a_{L} + \frac{{b_{L}} }{{\left( {n - a_{L}}
\right)^{2}}} + \frac{{c_{L}} }{{\left( {n - a_{L}}  \right)^{4}}} + &\nonumber\\
& + \frac{{d_{L}} }{{\left( {n - a_{L}}  \right)^{6}}} + \frac{{e_{L}
}}{{\left( {n - a_{L}}  \right)^{8}}}...&
\end{align}

%========================================================================================

\begin{table*}
\caption{Quantum defects of alkali-metal Rydberg states}
\begin{tabular*}{\textwidth}{@{\extracolsep{\fill}}p{41pt}|p{16pt}|p{69pt}|p{67pt}|p{72pt}|p{71pt}|p{72pt}|p{71pt}}
\hline
\hline
&&
\textit{S}$_{1/2}$& 
\textit{P}$_{1/2}$& 
\textit{P}$_{3/2}$& 
\textit{D}$_{3/2}$& 
\textit{D}$_{5/2}$& 
\textit{F}$_{5/2}$ \par \textit{F}$_{7/2}$ \\
\hline
\raisebox{-6.00ex}[0cm][0cm]{Li~\cite{Li1996}}& 
\textit{a}& 
0.39951183& 
0.04716876& 
& 
0.00194211& 
& 
0.00030862 \\
\cline{2-8} 
 & 
\textit{b}& 
0.02824560& 
-0.02398188& 
& 
-0.00376875& 
& 
-0.00099057 \\
\cline{2-8} 
 & 
\textit{c}& 
0.02082123& 
0.01548488& 
& 
-0.01563348& 
& 
-0.00739661 \\
\cline{2-8} 
 & 
\textit{d}& 
-0.09793152& 
-0.16065777& 
& 
0.10335313& 
& 
 \\
\cline{2-8} 
 & 
\textit{e}& 
0.14782202& 
0.33704280& 
& 
& 
& 
 \\
\hline
\raisebox{-4.50ex}[0cm][0cm]{Na~\cite{Na1995}}& 
\textit{a}& 
1.34796938(11)& 
0.85544502(15)& 
0.85462615(12)& 
0.014909286(97)& 
0.01492422(16)& 
0.001453* \\
\cline{2-8} 
 & 
\textit{b}& 
0.060989(16)& 
0.112067(86)& 
0.112344(67)& 
-0.042506(35)& 
-0.042585(43)& 
0.017312* \\
\cline{2-8} 
 & 
\textit{c}& 
0.019674(17)& 
0.0479(13)& 
0.0497(10)& 
0.00840(31)& 
0.00840(39)& 
-0.7809* \\
\cline{2-8} 
 & 
\textit{d}& 
-0.001045(354)& 
0.0457(43)& 
0.0406(34)& 
& 
& 
7.021* \\
\hline
\raisebox{-6.00ex}[0cm][0cm]{K~\cite{Def1983}}& 
\textit{a}& 
\textsf{2.1801985}& 
\textsf{1.713892}& 
\textsf{1.710848}& 
\textsf{0.276970}& 
\textsf{0.2771580}& 
\textsf{0.010098} \\
\cline{2-8} 
 & 
\textit{b}& 
\textsf{0.13558}& 
\textsf{0.233294}& 
\textsf{0.235437}& 
\textsf{-1.024911}& 
\textsf{-1.025635}& 
\textsf{-0.100224} \\
\cline{2-8} 
 & 
\textit{c}& 
\textsf{0.0759}& 
\textsf{0.16137}& 
\textsf{0.11551}& 
\textsf{-0.709174}& 
\textsf{-0.59201}& 
\textsf{1.56334} \\
\cline{2-8} 
 & 
\textit{d}& 
\textsf{0.117}& 
\textsf{0.5345}& 
\textsf{1.1015}& 
\textsf{11.839}& 
\textsf{10.0053}& 
\textsf{-12.6851} \\
\cline{2-8} 
 & 
\textit{e}& 
\textsf{-0.206}& 
\textsf{-0.234}& 
\textsf{-2.0356}& 
\textsf{-26.689}& 
\textsf{-19.0244}& 
\textsf{} \\
\hline
\raisebox{-4.50ex}[0cm][0cm]{Rb~\cite{Rb2003,RbF}}& 
\textit{a} \par \textit{}& 
\textsf{3.1311804(10)}& 
\textsf{2.6548849(10)}& 
\textsf{2.6416737(10)}& 
\textsf{1.34809171(40)}& 
\textsf{1.34646572(30)}& 
0.0165192(9) \textsf{(}\textsf{\textit{F}}$_{5/2}$\textsf{)} \\
\cline{2-8} 
 & 
\textit{}& 
\textsf{}& 
\textsf{}& 
\textsf{}& 
\textsf{}& 
\textsf{}& 
\textsf{0.0165437(7) (}\textsf{\textit{F}}$_{7/2}$\textsf{)} \par  \\
\cline{2-8} 
 & 
\textit{}& 
\textsf{0.1784(6)}& 
\textsf{0.2900(6)}& 
\textsf{0.2950(7)}& 
\textsf{-0.60286(26)}& 
\textsf{-0.59600(18)}& 
\textsf{-0.085(9) (}\textsf{\textit{F}}$_{5/2}$\textsf{)} \\
\cline{2-8} 
 & 
\textit{b}& 
\textsf{}& 
\textsf{}& 
\textsf{}& 
\textsf{}& 
\textsf{}& 
\textsf{-0.086(7) (}\textsf{\textit{F}}$_{7/2}$\textsf{)} \\
\hline
\raisebox{-6.00ex}[0cm][0cm]{Cs~\cite{Cs1987}}& 
\textit{a}& 
\textsf{4.04935665(38)}& 
\textsf{3.59158950(58)}& 
\textsf{3.5589599**}& 
\textsf{2.4754562**}& 
\textsf{2.46631524(63)}& 
\textsf{0.03341424(96)} \\
\cline{2-8} 
 & 
\textit{b}& 
\textsf{0.2377037}& 
\textsf{0.360926}& 
\textsf{0.392469**}& 
\textsf{0.009320**}& 
\textsf{0.013577}& 
\textsf{-0.198674} \\
\cline{2-8} 
 & 
\textit{c}& 
\textsf{0.255401}& 
\textsf{0.41905}& 
\textsf{-0.67431**}& 
\textsf{-0.43498**}& 
\textsf{-0.37457}& 
\textsf{0.28953} \\
\cline{2-8} 
 & 
\textit{d}& 
\textsf{0.00378}& 
\textsf{0.64388}& 
\textsf{22.3531**}& 
\textsf{-0.76358**}& 
\textsf{-2.1867}& 
\textsf{-0.2601} \\
\cline{2-8} 
 & 
\textit{e}& 
\textsf{0.25486}& 
\textsf{1.45035}& 
\textsf{-92.289**}& 
\textsf{-18.0061**}& 
\textsf{-1.5532}& 
\textsf{} \\
\hline
\multicolumn{8}{p{482pt}}{*\cite{Def1983} \par **\cite{Cs1984}\textsf{}}  \\
\end{tabular*}
\end{table*}

%=========================================================================================

The difference between the fitting coefficients in Eqs.(\ref{eq3}) and (\ref{eq4}) is small
and lies within the measurement uncertainty~\cite{Def1983}. The most recent
experimental values of the modified Rydberg-Ritz coefficients~\cite{Cs1984}
available for alkali-metal Rydberg atoms are listed in Table~1 of Appendix.
The data are taken from Refs.~\cite{Li1996, Na1995, Def1983, Rb2003, RbF,
Cs1987,Cs1984}.

We have compared our calculations with our recent experimental data for
diffuse series in rubidium. Good agreement between experiment and theory is
observed.

For the calculation of the dipole matrix elements of bound-free and
free-free transitions, the expressions for the quantum defects of Rydberg
states, given by Eqs.(\ref{eq3}) and (\ref{eq4}), must be extrapolated to the continuum~\cite{BurgessSeaton}:

\begin{equation}
\label{eq5}
\begin{array}{l}
 \mu _{L} \left( {E} \right) = a_{L} + b_{L} \times \left( { - 2E} \right) +
\\
 + c_{L} \times \left( { - 2E} \right)^{2} + d_{L} \times \left( { - 2E}
\right)^{3}... \\
 \\
 \end{array}
\end{equation}

\noindent Here \textit{E} is energy of the continuum state. We have found that
calculations of the radial matrix elements for bound-free and free-free
transitions are sensitive to the way of extrapolation of the quantum defects
into continuum, especially in the regions of the Cooper minima. This was
also discussed earlier in Ref.~\cite{Bezuglov1999}. The most recent data for the
rubidium quantum defects~\cite{Rb2003,RbF} contain only two coefficients $a_{L}
$ and $b_{L} $, while for the other alkali-metal atoms up to five terms of
Eq.(\ref{eq4}) have been published~\cite{Li1996, Na1995, Def1983, Cs1987, Cs1984}.
However, our test for using the higher-order polynomial approximations gave
incorrect values of the quantum defects at large energies of the continuum
states, due to the well known Runge's phenomenon~\cite{Runge} of oscillation of
the interpolation function near the edge of the interpolation region, when the
higher-order polynomial approximation is used. Therefore, in our
calculations we applied a linear extrapolation of quantum defects to the
continuum with only the first two terms of Eq.~(\ref{eq5}).

\section{Dyachkov-Pankratov quasiclassical model}

\begin{widetext}

A quasiclassical model of Dyachkov and Pankratov was published in their
original papers~\cite{Dyachkov1994, DyachkovFree}. Here we summarize the main
formulas, which are used to calculate the relative matrix elements for
bound-bound, bound-free and free-free transitions between $\left| {E,L}
\right\rangle $ and $\left| {{E}',{L}'} \right\rangle $ states and
transition frequency $\omega = {E}' - E$, where${E}' > E$. The difference of
the quantum defects $\Delta \mu = \mu _{{L}'} \left( {{E}'} \right) - \mu
_{L} \left( {E} \right)$.

The Keppler motion of quasiclassical electron is determined by the mean
energy:

\begin{equation}
\label{eq6}
E_{c} = \frac{{E + {E}'}}{{2}}.
\end{equation}

\noindent The cases of $E_{c} < 0$ (finite mean orbit) and $E_{c} > 0$ (infinite mean
orbit) must be considered separately. If $E_{c} < 0$ (bound-bound
transitions and bound-free transitions to the continuum states with ${E}' <
\left| {E} \right|$), the parameters $\gamma ^{\ast} $, $\gamma $ and mean
quantum number $\nu _{c} $ are defined as:

\begin{align}
\label{eq7}
 &\gamma ^{\ast}  = \frac{{\omega} }{{\left| {E + {E}'} \right|^{3/2}}}, &\nonumber\\
 &\gamma = int\left[ {\gamma ^{\ast}  + \Delta \mu + 0.5} \right] - \Delta
\mu , &\nonumber\\
 &\nu _{C} = \left( {{{\gamma}  \mathord{\left/ {\vphantom {{\gamma}  {\omega
}}} \right. \kern-\nulldelimiterspace} {\omega} }} \right)^{1/3}. &
 \end{align}

\noindent Here $int[x]$ means integer part of \textit{x}. In the quasiclassical
model the dynamics of the electron is defined by the arithmetic mean orbital
momentum $l_{c} $ and eccentricity of the mean elliptic orbit of the
electron $\varepsilon _{c} $:

\begin{equation}
\label{eq8}
l_{c} = \frac{{L + {L}' + 1}}{{2}},\quad \varepsilon = \left( {1 -
\frac{{l_{c}^{2}} }{{\nu _{c}^{2}} }} \right)^{1/2}.
\end{equation}

\noindent The relative matrix element is then expressed as:

\begin{align}
\label{eq9}
&R_{rel} \left( {EL \to {E}'{L}'} \right) = \frac{{\nu _{c}^{2} \omega
^{2/3}}}{{0.4108}}\left[ {U_{\gamma}  \left( {\varepsilon \gamma}
\right)\cos\left( {\pi \Delta \mu}  \right) - V_{\gamma}  \left( {\varepsilon
\gamma}  \right)\sin\left( {\pi \Delta \mu}  \right)} \right], & \nonumber\\
 &U_{\gamma}  = {J}'_{\gamma}  \left( {\varepsilon \gamma}  \right) + \Delta
l\frac{{l_{c}} }{{\nu _{c} \varepsilon} }J_{\gamma}  \left( {\varepsilon
\gamma}  \right), & \nonumber\\
 &V_{\gamma}  = {E}'_{\gamma}  \left( {\varepsilon \gamma}  \right) + \Delta
l\frac{{l_{c}} }{{\nu _{c} \varepsilon} }\left[ {E_{\gamma}  \left(
{\varepsilon \gamma}  \right) - \frac{{1}}{{\pi \gamma} }} \right] +
\frac{{1 - \varepsilon} }{{\pi}}.&
\end{align}

\noindent Here $J_{\gamma}  $ and $E_{\gamma}  $ are the Anger and Weber functions,
respectively, ${J}'_{\gamma}  $ and ${E}'_{\gamma}  $ are their derivatives
with respect to argument.

If $E_{c} > 0$ (bound-free transitions with ${E}' > \left| {E} \right|$ and
free-free transitions), the relative matrix element is given by:

\begin{align}
\label{eq10}
&R_{rel} \left( {EL \to {E}'{L}'} \right) = \frac{{\eta _{c}^{2} \omega
^{2/3}}}{{0.4108}}\left[ {P_{\gamma}  \left( {\varepsilon \gamma}
\right)\cos\left( {\pi \Delta \mu}  \right) - Q_{\gamma}  \left( {\varepsilon
\gamma}  \right)\sin\left( {\pi \Delta \mu}  \right)} \right], &\nonumber\\
& P_{\gamma}  = - {g}'_{\gamma}  \left( {\varepsilon \gamma}  \right) +
\Delta l\frac{{l_{c}} }{{\eta _{c} \varepsilon} }g_{\gamma}  \left(
{\varepsilon \gamma}  \right), &\nonumber\\
 &Q_{\gamma}  = {h}'_{\gamma}  \left( {\varepsilon \gamma}  \right) + \Delta
l\frac{{l_{c}} }{{\eta _{c} \varepsilon} }\left[ {h_{\gamma}  \left(
{\varepsilon \gamma}  \right) - \frac{{1}}{{\pi \gamma} }} \right] +
\frac{{\varepsilon - 1}}{{\pi} }. &
\end{align}

\noindent Here $\eta _{c} = {{1} \mathord{\left/ {\vphantom {{1} {\sqrt {2E_{c}} } }}
\right. \kern-\nulldelimiterspace} {\sqrt {2E_{c}} } }$, $\gamma = \eta
_{c}^{3} \omega $ and $\varepsilon = \left( 1 + \l_c^2/\eta_c^2  \right)^{1/2}$.

Functions $g_{\gamma}  $ and $h_{\gamma}  $ are expressed through the Hankel
$H^{\left( {1} \right)}$, Anger $J$ and modified Bessel $I$ functions:

\begin{align}
\label{eq11}
&g_{\gamma}  \left( {y} \right) = \frac{{1}}{{2}}iH_{i\gamma} ^{\left( {1}
\right)} \left( {iy} \right),&\nonumber\\
&h_{\gamma}  \left( {y} \right) = \frac{{1}}{{\sinh\left( {\pi \gamma}
\right)}}\left\{ {J_{i\gamma}  \left( { - iy} \right) -
\frac{{1}}{{2}}\exp\left( {\pi \gamma /2} \right)\left[ {I_{iy} \left(
{\gamma}  \right) + I_{ - iy} \left( {\gamma}  \right)} \right]} \right\}& 
\end{align}

\noindent In the case of $E_{c} \approx 0$ an asymptotic expression for relative
matrix element can be used:

\begin{equation}
\label{eq12}
R_{rel} \left( {EL \to {E}'{L}'} \right) = \frac{{4^{1/3}}}{{0.4108}}\left\{
{\left[ {S\left( {x} \right) + \beta S_{1} \left( {x} \right)}
\right]\cos\left( {\pi \Delta \mu}  \right) - \left[ {T\left( {x} \right) +
\beta T_{1} \left( {x} \right)} \right]\sin\left( {\pi \Delta \mu}  \right)}
\right\}
\end{equation}

\noindent Here $x = \left( {{{l_{c}^{3} \omega}  \mathord{\left/ {\vphantom
{{l_{c}^{3} \omega}  {2}}} \right. \kern-\nulldelimiterspace} {2}}}
\right)^{2/3}$, $\beta = 2E_{c} \left( {{{2} \mathord{\left/ {\vphantom {{2}
{\omega} }} \right. \kern-\nulldelimiterspace} {\omega} }} \right)^{2/3}$,
and functions $S\left( {x} \right)$, $S_{1} \left( {x} \right)$, $T\left(
{x} \right)$, $T_{1} \left( {x} \right)$ are expressed through the Airy
$Ai\left( {x} \right)$ and $Bi\left( {x} \right)$ functions, their
derivatives and hypergeometric function ${}_{1}F^{2}$:

\begin{align}
\label{eq13}
& S\left( {x} \right) = \Delta l \cdot x^{1/2} \cdot Ai\left( {x} \right) -
A{i}'\left( {x} \right), &\nonumber\\
&T\left( {x} \right) = \Delta l \cdot x^{1/2} \cdot Gi\left( {x} \right) -
G{i}'\left( {x} \right) + \frac{{x}}{{2\pi} }, &\nonumber\\
&S_{1} \left( {x} \right) = \frac{{1}}{{10}}\left( {1 - 6\Delta l \cdot
x^{3/2} + 4x^{3}} \right)Ai\left( {x} \right) + \frac{{2}}{{5}}x\left( {1 -
\Delta l \cdot x^{3/2}} \right)A{i}'\left( {x} \right), &\nonumber\\
&T_{1} \left( {x} \right) = \frac{{1}}{{10}}\left( {1 - 6\Delta l \cdot
x^{3/2} + 4x^{3}} \right)Gi\left( {x} \right) + \frac{{2}}{{5}}x\left( {1 -
\Delta l \cdot x^{3/2}} \right)G{i}'\left( {x} \right) + \frac{{9}}{{20\pi
}}\Delta l \cdot x^{1/2} - \frac{{21}}{{40\pi} }x^{2}, &\nonumber\\
&Gi\left( {x} \right) = \frac{{1}}{{3}}Bi\left( {x} \right) -
\frac{{x^{2}}}{{2\pi} }{}_{1}F^{2}\left(
{1;\frac{{4}}{{3}},\frac{{5}}{{3}};\frac{{x^{3}}}{{3}}} \right). &\nonumber\\
&G{i}'\left( {x} \right) = \frac{{Bi\left( {x} \right)}}{{3}} - 
\frac{{x}}{{\pi} }{}_{1}F^{2}\left(
{1;\frac{{4}}{{3}},\frac{{5}}{{3}};\frac{{x^{3}}}{{9}}} \right) - 
\frac{{3x^{4}}}{{40\pi} }{}_{1}F^{2}\left(
{2;\frac{{7}}{{3}},\frac{{8}}{{3}};\frac{{x^{3}}}{{9}}} \right) +
\frac{{x}}{{2\pi} }. &
\end{align}

\noindent We note that in the original paper~\cite{Dyachkov1994} there were misprints in
the last two terms in the expression for $T_{1} \left( {x} \right)$. Here we
present the corrected formula kindly provided by L.~G.~Dyachkov
\cite{DyachkovPrivate}.

To simplify numerical calculations with generic mathematical codes and
software, functions $J_{\gamma}  $, $E_{\gamma}  $, $g_{\gamma}  $
,$h_{\gamma}  $ and their derivatives can be expressed via more commonly
used regularized hypergeometric functions ${}_{1}\tilde {F}^{2}$,
hypergeometric functions ${}_{1}F^{2}$, Hankel \textit{H} and modified
Bessel $I$ functions \cite{Abramowitz}:

\begin{align}
\label{eq14}
& J_{\gamma}  \left( {z} \right) = \frac{{1}}{{2}}z\sin\left( {\frac{{\pi
\gamma} }{{2}}} \right){}_{1}\tilde {F}^{2}\left[ {1;\,\frac{{1}}{{2}}\left(
{3 - \gamma}  \right),\frac{{1}}{{2}}\left( {3 + \gamma}  \right);\; -
\frac{{z^{2}}}{{4}}} \right] + \cos\left( {\frac{{\pi \gamma} }{{2}}}
\right){}_{1}\tilde {F}^{2}\left[ {1;\,1 - \frac{{\gamma} }{{2}},\;1 +
\frac{{\gamma} }{{2}};\, - \frac{{z^{2}}}{{4}}} \right], &\nonumber \\
& E_{\gamma}  \left( {z} \right) = - \frac{{1}}{{2}}z\cos\left( {\frac{{\pi
\gamma} }{{2}}} \right){}_{1}\tilde {F}^{2}\left[ {1;\,\frac{{1}}{{2}}\left(
{3 - \gamma}  \right),\frac{{1}}{{2}}\left( {3 + \gamma}  \right);\; -
\frac{{z^{2}}}{{4}}} \right] + \sin\left( {\frac{{\pi \gamma} }{{2}}}
\right){}_{1}\tilde {F}^{2}\left[ {1;\,1 - \frac{{\gamma} }{{2}},\;1 +
\frac{{\gamma} }{{2}};\, - \frac{{z^{2}}}{{4}}} \right], &\nonumber\\
 &{J}'_{\gamma}  \left( {z} \right) = - \frac{{1}}{{2}}z\cos\left( {\frac{{\pi
\gamma} }{{2}}} \right){}_{1}\tilde {F}^{2}\left[ {2;\;2 - \frac{{\gamma
}}{{2}},\;2 + \frac{{\gamma} }{{2}};\; - \frac{{z^{2}}}{{4}}} \right] + &\nonumber\\
& \quad \quad + \frac{{1}}{{2}}\sin\left( {\frac{{\pi \gamma} }{{2}}}
\right){}_{1}\tilde {F}^{2}\left[ {1;\;\frac{{3 - \gamma} }{{2}},\;\frac{{3
+ \gamma} }{{2}};\; - \frac{{z^{2}}}{{4}}} \right] -
\frac{{z^{2}}}{{4}}\sin\left( {\frac{{\pi \gamma} }{{2}}} \right){}_{1}\tilde
{F}^{2}\left[ {1;\;1 + \frac{{3 - \gamma} }{{2}},1 + \frac{{3 + \gamma
}}{{2}}; - \frac{{z^{2}}}{{4}}} \right], &\nonumber\\
 &E_{\gamma}  \left( {z} \right) = - \frac{{1}}{{2}}z\sin\left( {\frac{{\pi
\gamma} }{{2}}} \right){}_{1}\tilde {F}^{2}\left[ {2;\;2 - \frac{{\gamma
}}{{2}},\;2 + \frac{{\gamma} }{{2}};\; - \frac{{z^{2}}}{{4}}} \right] - &\nonumber\\
 &\quad \quad - \frac{{1}}{{2}}\cos\left( {\frac{{\pi \gamma} }{{2}}}
\right){}_{1}\tilde {F}^{2}\left[ {1;\;\frac{{3 - \gamma} }{{2}},\;\frac{{3
+ \gamma} }{{2}};\; - \frac{{z^{2}}}{{4}}} \right] +
\frac{{z^{2}}}{{4}}\cos\left( {\frac{{\pi \gamma} }{{2}}} \right){}_{1}\tilde
{F}^{2}\left[ {1;\;1 + \frac{{3 - \gamma} }{{2}},1 + \frac{{3 + \gamma
}}{{2}}; - \frac{{z^{2}}}{{4}}} \right]. &\nonumber \\
 &{g}'_{\gamma}  \left( {z} \right) = \frac{{1}}{{4}}\left[ { - H_{ - 1 +
i\gamma} ^{\left( {1} \right)} \left( {iz} \right) + H_{1 + i\gamma
}^{\left( {1} \right)} \left( {iz} \right)} \right], &\nonumber\\
& {h}'_{\gamma}  \left( {z} \right) = \cosh\left( {\pi \gamma}  \right)\left\{
{ - \frac{{1}}{{4}}\exp\left( {\frac{{\pi \gamma} }{{2}}} \right)\left[ {I_{
- 1 - i\gamma}  \left( {z} \right) + I_{1 - i\gamma}  \left( {z} \right) +
I_{ - 1 + i\gamma}  \left( {z} \right) + I_{1 + i\gamma}  \left( {z}
\right)} \right]} \right\} + &\nonumber\\
 &+ \frac{{1}}{{2}}z\,\cosh\left( {\frac{{\pi \gamma} }{{2}}}
\right){}_{1}F^{2}\left( {2;2 - \frac{{i\gamma} }{{2}},2 + \frac{{i\gamma
}}{{2}};\frac{{z^{2}}}{{4}}} \right) + \frac{{1}}{{2}}\sinh\left( {\frac{{\pi
\gamma} }{{2}}} \right){}_{1}F^{2}\left( {1;\frac{{3 - i\gamma
}}{{2}},\frac{{3 + i\gamma} }{{2}};\frac{{z^{2}}}{{4}}} \right) + &\nonumber\\
 &+ \frac{{z^{2}}}{{4}}{}_{1}F^{2}\left( {1;1 + \frac{{3 - i\gamma} }{{2}},1
+ \frac{{3 + i\gamma} }{{2}};\frac{{z^{2}}}{{4}}} \right). &
 \end{align}

\end{widetext}

\bibliography{bib}

\end{document}